\documentclass[11pt]{article}
\usepackage{moriond,epsfig}

\usepackage{amsmath}
\usepackage{amsfonts}
\usepackage{cases}

\def\Journal#1#2#3#4{{#1} {\bf #2}, #3 (#4)}

\def\PRD{{\em Phys. Rev.} D}

\def\be{\begin{equation}}
\def\ee{\end{equation}}
\def\bea{\begin{eqnarray}}
\def\eea{\end{eqnarray}}

\begin{document}
\vspace*{4cm}
\title{ELECTROSTATIC ACCELEROMETER WITH BIAS REJECTION FOR DEEP SPACE GRAVITATION TESTS}

\author{B. LENOIR$^*$, B. CHRISTOPHE$^*$, S. REYNAUD$^\dag$}

\address{$^*$Onera -- The French Aerospace Lab\\
29 avenue de la Division Leclerc, F-92322 Ch\^atillon, France\\
$^\dag$Laboratoire Kastler Brossel (LKB), ENS, UPMC, CNRS\\
Campus Jussieu, F-75252 Paris Cedex 05, France}

\maketitle

The trajectory of an interplanetary spacecraft can be used to test gravitation in the Solar System. Its determination relies on radio tracking and is limited by the uncertainty on the spacecraft non-gravitational acceleration. But the addition of an accelerometer on board provides another observable which measures the departure of the probe from geodesic motion.

Such a concept has been proposed for the OSS mission~\cite{christophe2011oss} which embarks the Gravity Advanced Package~\cite{lenoir2011electrostatic}. This instrument, which is the focus of this article, is designed to make unbiased acceleration measurements~\cite{lenoir2011unbiased}. This proposal is in line with the Roadmap for Fundamental Physics in Space issued by ESA in 2010~\cite{esa2010roadmap}. Indeed, there exist theoretical as well as experimental motivations to test gravitation in the Solar System. The fact that General Relativity is a classical theory while the three other fundamental interactions have a quantum description suggests that it is not the final theory for gravitation. From the experimental point of view, the existence of ``dark matter'' and ``dark energy'' may be interpreted as the inability of General Relativity to describe gravitation at cosmic scales~\cite{aguirre2001astrophysical}. And in the Solar System, the anomalous Pioneer signal~\cite{anderson2002study,levy2009pioneer} may be an experimental artifact as well as a hint of considerable importance for fundamental physics~\cite{jaekel2005gravity}.

\section{Presentation of the instrument}

The Gravity Advanced Package is made of two subsystems: MicroSTAR, an electrostatic accelerometer which inherits mature technology developed at Onera, and the Bias Rejection System used to rotate MicroSTAR with respect to the spacecraft around one axis. MicroSTAR measures the non-gravitational acceleration of the satellite and the measurement noise is characterized by the following power spectrum density~\cite{lenoir2011electrostatic} (for a measurement range of $1.8 \times 10^{-4}$ m.s$^{-2}$):
\begin{equation}
  S_{n}(f) = \left(5.7 \times 10^{-11} \ \mathrm{m}.\mathrm{s}^{-2}.\mathrm{Hz}^{-1/2}\right)^2 \times \left[ 1 + \left(\frac{f}{4.2 \ \mathrm{mHz}}\right)^{-1} + \left(\frac{f}{0.27 \ \mathrm{Hz}}\right)^{4}\right].
  \label{eq:acc_noise}
\end{equation}

Thanks to the Bias Rejection System, which rotates the accelerometer of a monitored angle called $\theta$, the quantities measured along the orthogonal axis $y$ and $z$ of the accelerometer are (assuming that there is no quadratic terms and the gain of the instrument is perfectly known)
\begin{subequations}
  \begin{numcases}{}
    m_y = \left[\cos(\theta) a_Y  + \sin(\theta) a_Z \right] + b_y + n_y \\
    m_z = \left[-\sin(\theta) a_Y + \cos(\theta) a_Z \right] + b_z + n_z
  \end{numcases}
  \label{eq:measurement}
\end{subequations}
with $a_\nu$ ($\nu\in \{Y;Z\}$) the components of the acceleration in the reference frame of the spacecraft, $b_y$ and $b_z$ the bias of MicroSTAR on each axis, and $n_y$ and $n_z$ the measurement noise.

\section{Signal processing}

Assuming that $N$ measurements are made with a time step $\delta t$, there are $4N$ unknowns in equations \eqref{eq:measurement} and only $2N$ measurements. Calling $\mathbf{x}$ the column vectors whose components are the values of $x$ at each sampling time and using the linearity of the equations, it is however possible to retrieve from the measurements the values of the projection of $\mathbf{a_Y}$ and $\mathbf{a_Z}$ on a vector subspace defined by the columns of the matrix $V_a \in \mathcal{M}_{N,p_a}$ ($p_a < N$). If
\begin{equation}
  {V_a}' \Lambda_\nu \mathbf{b_\kappa} = 0, \ \mathrm{with} \ \nu\in\{c;s\} \ \mathrm{and} \ \kappa\in\{\mathbf{y};\mathbf{z}\}.
  \label{eq:condition_demodulation}
\end{equation}
where $\Lambda_c = \mathrm{diag}[\cos(\theta_k)]$ and $\Lambda_s = \mathrm{diag}[\sin(\theta_k)]$ ($k\in||1;N||$), then
\begin{subequations}
  \begin{numcases}{}
    V_a' \mathbf{a_Y} = V_a' \Lambda_c \mathbf{m_y} - V_a' \Lambda_s \mathbf{m_z} \\
    V_a' \mathbf{a_Z} = V_a' \Lambda_s \mathbf{m_y} + V_a' \Lambda_c \mathbf{m_z}
  \end{numcases}
  \label{eq:a_simplified}
\end{subequations}
Assuming that the bias on each axis also belongs to a subspace defined by the matrix $V_b$, is is possible to design pattern for the rejection angle $\theta$ which fulfills conditions \eqref{eq:condition_demodulation}. This signal has a period called $\tau$.

In addition to retrieving the unbiased non-gravitational acceleration of the spacecraft, this method allows characterizing the uncertainty on the demodulated quantities $V_a' \mathbf{a_Y}$ and $V_a' \mathbf{a_Z}$, given MicroSTAR noise power spectrum density (cf. eq. \eqref{eq:acc_noise}). Assuming that $V_a \in \mathcal{M}_{N,1}(\mathbb{R})$ and $|V_a| = 1$, the precision on the quantities $V_a' \mathbf{a_Y}$ and $V_a' \mathbf{a_Z}$ is given by~\cite{lenoir2011unbiased}
\begin{equation}
  \int_{-\frac{1}{2\delta t}}^{\frac{1}{2\delta t}} S_n(f)  \frac{\left|\mathcal{F}_{\delta t}\{\Lambda_c V_a\}(f)\right|^2 + \left|\mathcal{F}_{\delta t}\{\Lambda_s V_a\}(f)\right|^2}{\delta t^2}  df \approx \frac{1}{\tau} S_n\left(\frac{1}{\tau}\right)
  \label{eq:uncertainty}
\end{equation}
where $\mathcal{F}_{\delta t}$ is the Discrete Time Fourier Transform. One has to notice that the noise is selected, as expected, at the modulation frequency $1/\tau$.

\section{Experimental validation}

This demodulation scheme will be validated experimentally at Onera using a pendulum. A control loop allows controlling its inclination to the $10^{-9}$ rad level. It is possible to incline it at a known angle in order to simulate an external acceleration. On this pendulum, an accelerometer is mounted on a rotating stage. There are two goals for this experiment :

-- Validate the demodulation scheme by showing that it properly separates the bias from the signal of interest allowing to make unbiased acceleration measurements.

-- Verify the value of the uncertainty on $V_a' \mathbf{a_Y}$ and $V_a' \mathbf{a_Z}$ predicted by equation \eqref{eq:uncertainty}.

\end{document}